\documentclass[twocolumn,twoside,slac_two]{revtex4}
\usepackage{graphicx}
\usepackage{fancyhdr}
\begin{document}

\title{ Direct CP Violation in Charmless $B$ Decays}

%

\author{Y. Chao}
\affiliation{National Taiwan University,  No.1, Sec. 4, Roosevelt Road,
Taipei, Taiwan 106}

\begin{abstract}
We report the updated results of direct $CP$ violation measurements from
Belle and BaBar for the following modes: $B \to h^+h^-/h^\pm\pi^0$,
$B \to \eta{^(}'{^)} h^\pm$ and $B \to K^\pm\pi^\pm\pi^\mp$, where $h$ stands
for a charged kaon or pion. The data used in these
studies are up to 386 million $B \bar B$ pairs for Belle and 232 million
$B \bar B$ pairs for BaBar. The current significant measurements include:
$\mathcal A_{CP}(B \to K^+ \pi^-)=-0.108\pm0.016$,
$\mathcal A_{CP}(B \to \rho^0(770) K^+)=0.31^{+0.11}_{-0.10}$, and
$\mathcal A_{CP}(B \to \eta K^+)=-0.33\pm0.12$~\footnote{
http://www.slac.stanford.edu/xorg/hfag}.

\end{abstract}

\maketitle

\thispagestyle{fancy}


\section{Introduction}
In the Standard Model (SM) $CP$ violation arises via the interference of at
least two diagrams with comparable amplitudes but different $CP$ conserving
and violating phases. Mixing induced $CP$ violation in the $B$ sector has been
established in $b\to c\bar{c} s$ transitions \cite{2phi1,2beta}. In the SM,
direct $CP$ violation is expected to be sizable in the $B$ meson system
\cite{BSS}. Charmless $B$ decays provide a  rich sample to understand $B$
decay dynamics and to search for $CP$ violation.
The first experimental evidence for direct $CP$
violation was shown by Belle  for the decay mode $B^0\to \pi^+\pi^-$
\cite{PIPI}. This result suggests large interference between tree and penguin
diagrams and the existence of final state interactions \cite{FSI}.

The three-body charmless hadronic $B$ mesons decays can provide new
possibilities for $CP$ violation searches. In spite of measuring the
difference between $B$ and $\bar{B}$ decay rates, one can also measure
the difference in relative phase between two quasi-two-body amplitudes,
which are often dominated in three-body $B$ decays. This can be achieved via
amplitude (Dalitz) analysis. As currently direct $CP$ violation has been
observed only in decays of neutral $K$ mesons~\cite{dcpv-K0} and recently
in neutral $B$ meson decays~\cite{dcpv-B0}. Large direct $CP$ violation is
expected in charged $B$ decays to some quasi-two-body charmless hadronic
modes~\cite{beneke-neubert}.


The partial rate $CP$ violating asymmetry is defined as:
\begin{eqnarray}
\mathcal A_{CP}=\frac{N(\overline B \to \overline f)-N(B \to f)}
{N(\overline B \to \overline f)+N(B \to f)},
\end{eqnarray}
where $N(\overline B \to \overline f)$ is the yield for the
$\overline{B}$ decays and $N(B \to f)$ denotes that of the
charge-conjugate mode.

We report the updated direct $CP$ violating measurements from both
Belle and BaBar. The data used in these study are up to 386 milion
$B \bar B$ pairs collected at Belle~\cite{Belle} detector and 274 milion
$B \bar B$ pairs at BaBar. The Belle detector is located at the KEKB $e^+e^-$
asymmetric-energy (3.5 on 8~GeV) collider~\cite{KEKB} and BaBar
detector~\cite{babar} is at the SLAC PEP-II asymmetric-energy (3.1 on 9~GeV)
$e^+e^-$ storage ring~\cite{pep2}. Both of them are operating at the
$\Upsilon(4S)$ resonance.

\section{Event Reconstruction}
Charged tracks are required to have momenta transverse to the beam
greater than 0.1~GeV/$c$ and to be consistent with originating from
the interaction point (IP). For particle identification of charged tracks,
a combined likelihood of information from $dE/dx$ of drift chamber,
time-of-flight and aerogel Cherenkov counter is used by Belle while an
associated Cherenkov-angle from the ring image counter (DIRC) is used by BaBar.
For some of the modes described below, charged tracks that are positively
identified as electrons or protons are excluded. Candidate $\pi^0$ mesons
are selected by requiring the two-photon invariant mass to be in the
$2.5\sim 3.5\sigma$ mass window. The momentum vector of each photon is
then readjusted to constrain the mass of the photon pair to the nominal
$\pi^0$ mass. Candidate $\eta$ mesons are reconstructed by combining a
$\pi^0$ with at least 250 MeV/$c$ laboratory momentum with a pair of
oppositely charged tracks that originate from the interaction point (IP).
For Belle, the following requirements are made on the invariant mass of
the $\eta$ candidates: 516 MeV/$c^2 < M_{\gamma\gamma} < 569$ MeV/$c^2$ for
$\eta \to \gamma \gamma$ and 539 MeV/$c^2 < M_{3\pi} <556$ MeV/$c^2$ for
$\eta \to 3\pi$'s. As to BaBar, the following cuts are applied:
490 MeV/$c^2 < M_{\gamma\gamma} < 600$ MeV/$c^2$ for $\eta_{\gamma \gamma}$
and 520 MeV/$c^2 < M_{3\pi} <570$ MeV/$c^2$ for $\eta_{3\pi}$.
After the selection of each candidate, the $\eta$ mass constraint is
implemented by readjusting the momentum vectors of the daughter particles.
The $\eta'$ mesons are reconstructed via two decay chains:
$\eta \pi \pi$ (with $\eta \to \gamma \gamma)$ and $\eta' \to \rho \gamma$.
In addition, we require the following. All photons are required to have
an energy of at least 50 MeV, photons from $\eta'$ in $\eta' \to \rho \gamma$
of at least 100 MeV. The transverse momenta of $\pi^\pm$ for $\rho^0$
candidates have to be greater than 200 MeV/$c$.
$K^0_S$ candidates are reconstructed from pairs of oppositely-charged tracks
with an invariant mass ($M_{\pi\pi}$) between 480 and 516 MeV/$c^2$.
Each candidate must have a displaced vertex with a flight direction
consistent with that of a $K^0_S$ originating from the IP.

$B$ meson candidates are identified using two kinematic variables:
the energy difference $\Delta E = E_B  - E_{\mbox{\scriptsize beam}},$
and the beam constrained mass
$M_{\rm bc} =  \sqrt{E^2_{\mbox{\scriptsize beam}} - P_B^2},$
$E_{\mbox{\scriptsize beam}}$ is the run-dependent beam energy in the
$\Upsilon(4S)$ rest frame and  is determined  from
$B\to D^{(*)}\pi$ events, and $P_{B}$ and $E_B$ are the momentum and
energy of  the $B$ candidate in the $\Upsilon(4S)$ rest frame.
The resolutions on $M_{\rm bc}$ and $\Delta E$ are about 3 MeV/$c^2$ and
20--30 MeV, respectively. Events with $M_{\rm bc} >5.2$ GeV/$c^2$ and
$|\Delta E|<0.3$ GeV are selected for the analysis.

\section{Background reduction}
The dominant background comes from the $e^+e^-\rightarrow q\bar{q}$ continuum,
where $q= u, d, s$ or $c$. To distinguish signal from the jet-like continuum
background,  event shape variables and $B$ flavor tagging information
are employed. We combine information of correlated shape variables into a
Fisher discriminant \cite{fisher} and compute the likelihood as a product of
probabilities of this discriminant and $\cos\theta_B$, where $\theta_B$ is
the angle between the $B$ flight direction and the beam direction
in the $\Upsilon(4S)$ rest frame. A likelihood ratio,
${\cal LR} = {\cal L}_s/({\cal L}_s + {\cal L}_{q \bar{q}})$,
is formed from signal
(${\cal L}_s$) and background (${\cal L}_{q \bar{q}}$) likelihoods, obtained
using events from the signal Monte Carlo (MC) and from data with
$M_{\rm bc}< 5.26$ GeV/$c^2$, respectively.  Additional
background discrimination is provided by $B$ flavor tagging. An event that
contains a lepton (high quality tagging) is more likely to be a
$B \overline B$ event so a looser $\cal LR$ requirement can be applied.
We divide the data into six sub-samples based on the quality of flavor
tagging \cite{tagging}. Continuum suppression is achieved by applying a
mode dependent requirement on $\cal LR$ for events in each sub-sample
according to $N_s^{\rm exp}/\sqrt{N_s^{\rm exp}+N_{q\bar{q}}^{\rm exp}}$,
where $N_s^{\rm exp}$ is the expected signal  from MC and
$N_{q\bar{q}}^{\rm exp}$ denotes the number of background events estimated
from data.

\section{Signal Extraction}

\subsection{Maximum-likelihood Method}
The signal yields and partial rate asymmetries are obtained using an extended
unbinned maximum-likelihood (ML) fit with input variables $M_{\rm bc}$ and
$\Delta E$. The likelihood is defined as:
\begin{eqnarray}
\mathcal{L} & = &\rm{e}^{-\sum_j N_j}
\times \prod_i (\sum_j N_j \mathcal{P}_j) \;\;\; \mbox{and} \\
\mathcal{P}_j & = &\frac{1}{2}[1- q_i \cdot \cal A_{CP}{}_j ]
P_j(M_{{\rm bc}i}, \Delta E_i),
\end{eqnarray}
where $i$ is the identifier of the $i$-th event, $P(M_{\rm bc}, \Delta E)$
is the two-dimensional probability density function (PDF) in $M_{\rm bc}$
and $\Delta E$, $q$ indicates the $B$ meson flavor, $B^+(q=+1)$ or
$B^- (q=-1)$, $N_j$ is the number of events for the category $j$, which
corresponds to either signal, $q\bar{q}$ continuum, a reflection due to
$K$-$\pi$ misidentification, or background from other charmless $B$ decays.
For the neutral $B$ mode, $\mathcal{P}_j$ in the equation above is simply
$P_j(M_{{\rm bc}i}, \Delta E_i)$ and there is no reflection component.

\subsection{Dalitz Analysis}
As to the three-body direct $CP$ violation study, an unbinned maximum
likelihood fit method is also used. For the study performed by Belle,
the distribution of background events is parametrized by an empirical
function with 11 parameters\,\cite{khh-dalitz-belle}.
As found in the previous paper by Belle~\cite{khh-dalitz-belle},
the three-body $B^+ \to\ K\pi\pi^+$ amplitude is well-described by
a coherent sum of $K^*(892)^0\pi^+$, $K^*_0(1430)^0\pi^+$,
$\rho(770)^0K^+$, $f_0(980)K^+$, $f_X(1300)K^+$ and
$\chi_c K^+$ quasi-two-body channels and a non-resonant amplitude.
In order to describe an excess of signal events at
$M(\pi\pi)\simeq 1.3$~GeV/$c^2$ a $f_X(1300)K^+$ channel was introduced.
The best fit is achieved assuming $f_X(1300)$ is a scalar state;
the mass and width determined from the fit (see below) are consistent
with those for $f_0(1370)$~\cite{PDG}. Each quasi-two-body amplitude includes
a Breit-Wigner function, a $B$ decay form-factor parametrized in a
single-pole approximation, a Blatt-Weisskopf factor~\cite{blatt-weisskopf}
for the intermediate resonance decay, and a function that describes angular
correlations between final state particles. This is multiplied by a factor
of $ae^{i\delta}$ that describes the relative magnitude and phase of the
contribution.
The non-resonant amplitude is parametrized by an empirical function
${\cal A}_{\rm nr}(K\pi\pi^+)
      = a_1^{\rm nr}e^{-\alpha{s_{13}}}e^{i\delta^{\rm nr}_1}
      + a_2^{\rm nr}e^{-\alpha{s_{23}}}e^{i\delta^{\rm nr}_2},$
where $\alpha$, $a_i^{\rm nr}$ and $\delta_i^{\rm nr}$ are fit parameters,
$s_{13}\equiv M^2(K^+\pi^-)$, and $s_{23}\equiv M^2(\pi\pi)$.
In the analysis performed by Belle, a modified model that changing the
parameterization of the $f_0(980)$ line-shape from a Breit-Wigner function to
a Flatt\'e parameterization\,\cite{Flatte} and adding two more channels:
$\omega(782)K^+$ and $f_2(1270)K^+$ is used. For $CP$ violation studies
the amplitude for each quasi-two-body channel is modified from
$ae^{i\delta}$ to $ae^{i\delta}(1\pm be^{i\varphi})$, where the plus (minus)
sign corresponds to the $B^+$ ($B^-$) decay. With such a parameterization
the charge asymmetry, $\cal A_{CP}$, for a particular quasi-two-body
$B$$\to$$f$ channel is given by
\begin{equation}
A_{CP}(f)
      = \frac{N^--N^+}{N^-+N^+}
      = -\frac{2b\cos\varphi}{1+b^2}.
\label{eq:acp-dcpv}
\end{equation}
It is worth noting that in this parametrization zero relative phase
between $B^-$ and $B^+$ amplitudes is assumed.

\section{DCPV measurements}

\subsection{Two-body $B$ decays}
The first evidence of direct $CP$ violation in two-body $B$ decays in
$B^0 \to \pi^+ \pi^-$ was claimed by Belle~\cite{PIPI}. However,
this is not confirmed by the BaBar experiment~\cite{babar-pipi}.
There are both tree diagram and penguin diagram contributing to
$B \to K^+\pi^-$. Therefore, large $CP$ violation is expected in this decay
channel. In the summer of 2005, both Belle and BaBar reported
their updated measurements for the $K^+\pi^-$, $K^+\pi^0$ and $\pi^+\pi^0$
modes~\cite{b2kpi_2005}. In spite of the results from B factories, CDF
also report their measurements on the $K^+\pi^-$ mode and is included in the
world average shown in table~\ref{tab_b2kpi}. Note that the deviation
between $A_{CP}(B^0 \to K^+\pi^-)$ and $A_{CP}(B^+ \to K^+\pi^0)$ is
of $3.1\sigma$ for Belle only and the world average is now about $3.8\sigma$.

%

\begin{table}[h]
\begin{center}
\caption{Summary table of $A_{CP}(B\to hh)$.}
\begin{tabular}{lccc}
\hline
$\cal A_{CP}(\%)$ & BaBar & Belle & World Avg.\\
\hline
\hline
$K^+ \pi^-$ & $-13.3\pm3.0\pm.9$ &
$-11.3\pm2.2\pm.8$ & $-10.8\pm.17$ \\
\hline
$K^+\pi^0$ & $+6\pm6\pm1$ &
$+4\pm4\pm2$ & $+4\pm4$ \\
\hline
$\pi^+\pi^0$ & $-1\pm10\pm2$ &
$-2\pm8\pm1$ & $+1\pm6$ \\
\hline
\hline
\end{tabular}
\label{tab_b2kpi}
\end{center}
\end{table}

As to $A_{CP}(B\to \eta h)$, direct $CP$ violation is expected with
the interference between penguin processes and CKM suppressed tree process.
The $A_{CP}$ results are updated by both Belle and BaBar. A $2.9\sigma$
significance of $A_{CP}(B\to \eta K^+)$ is seen by Belle, but only
about $1 \sigma$ seen by BaBar.~\cite{b2etah_2005} While in the $B\to \eta' h$
decays, $B\to \eta' K^+$ is penguin dominated while $B\to \eta' \pi^+$ has
interference between tree and penguin. But there is no significant direct
$CP$ asymmetry seen by either experiments~\cite{b2etaph_2005}.
The summary of results are listed in table~\ref{tab_b2etah}.

\begin{table}[b]
\begin{center}
\caption{Summary table of $A_{CP}(B\to \eta^{(')}h)$.}
\begin{tabular}{lccc}
\hline
$\cal A_{CP}(\%)$ & BaBar & Belle & World Avg.\\
\hline
\hline
$B\to\eta K^+$ & $-20.\pm15.\pm1.$ &
$-55.\pm19.\pm4.$ & $-33.\pm12.$ \\
\hline
$B\to\eta\pi^+$ & $-13.\pm12.\pm1.$ &
$-10.\pm11.\pm2.$ & $-11.\pm8.$ \\
\hline
$B\to\eta\rho^+$ & $+2.\pm18.\pm2.$ &
$-17.\pm33.\pm2.$ & $-3.\pm16.$ \\
\hline
\hline
$B\to\eta' K^+$ & $+3.3\pm2.8\pm.5$ &
$+2.9\pm2.8\pm2.1$ & $+3.1\pm2.1$ \\
\hline
$B\to\eta' \pi^+$ & $+14.\pm16.\pm1.$ &
$+15.\pm38.\pm6.$ & $+14.\pm15.$ \\
\hline
\hline
\end{tabular}
\label{tab_b2etah}
\end{center}
\end{table}

%

\subsection{Three-body $B$ decays}
The direct $CP$ violation measurements in three-body $B$ decays are performed
with dalitz analysis by both Belle and BaBar in 2005~\cite{3body}. The
table~\ref{tab_3body} shows the direct $CP$ asymmetries for each quasi-two
body decay channels.

Among these quasi-two-body decay channels, Belle claimed the first evidence
of direct $CP$ asymmetry in $B \to \rho^0(770) K^\pm$ with $3.9 \sigma$
statistical significance. By comparing with null-asymmetry assumed toy MC,
they claim the significance would be about $3.7 \sigma$. Also, a
tensor-pseudoscalar mode $B \to f_2(1270) K^\pm$ was observed by Belle
with more than $6 \sigma$ significance and the charge asymmetry measurement
is given. This is not seen by BaBar and only an upper-limit on the branching
fraction is given.

\begin{table}[ht]
\begin{center}
\caption{Summary table of $A_{CP}$ in $B \to K^\pm\pi^\pm\pi^\mp$
three-body decays.}
\begin{tabular}{lcccc}
\hline\hline
$A_{CP}(\%)$ & BaBar & Belle & World Avg.\\
\hline
\hline
$K^\pm\pi^\pm\pi^\mp$ & $-1.3\pm3.7\pm1.1$ &
$+4.9\pm2.6\pm2.0$ & $+2.2\pm2.9$ \\
\hline
$K^*(892)\pi^\pm$ & $+7\pm8\pm7$ &
$-14.9\pm6.4\pm2.1$ & $-8.6\pm5.6$ \\
\hline
$K^*(1430)\pi^\pm$ & $-6.4\pm3.2^{+2.3}_{-2.6}$ &
$+7.6\pm3.8^{+2.8}_{-2.2}$ & $-0.2\pm2.9$ \\
\hline
$\rho(770)K^\pm$ & $+32\pm13^{+10}_{-8}$ &
$+30\pm11^{+11}_{-5}$ & $+31^{+11}_{-10}$ \\
\hline
$f_0(980)K^\pm$ & $+9\pm10^{+10}_{-6}$ &
$-7.7\pm6.5^{+4.6}_{-2.6}$ & $-2.6^{+6.8}_{-6.4}$ \\
\hline
$f_2(1270)K^\pm$ & --- &
$-59\pm22\pm4$ & $-59\pm22$ \\
\hline
$\chi_{c0} K^\pm$ &  --- &
$-6.5\pm20\pm2^{+29}_{-14}$ & --- \\
\hline
\hline
\end{tabular}
\label{tab_3body}
\end{center}
\end{table}

\section{Conclusions}
There are lots of fruitful experimental results since last FPCP meeting and
many improvements on the theoretical calculations. We now have observation of
direct $CP$ violation in $B \to K^+\pi^-$ mode and its sign and amplitude can
be better understood with next to leading order (NLO) calculations~\cite{NLO}.
However, the deviation between $A_{CP}(B \to K^+\pi^-)$ and
$A_{CP}(B \to K^+\pi^0)$ may need further clarifying theoretically~\cite{anom}.

Other than $B \to K^+\pi^-$, there are some hints or evidence of direct $CP$
asymmetry seen by Belle, but those are not yet confirmed by BaBar. Therefore,
more statistics are still needed for both B-factories to clarify the
situation.

\bigskip 
\begin{acknowledgments}
This proceeding for FPCP 2006 includes experimental results on charmless
$B$ meson decays measured by both Belle and BaBar.
The support for Belle experiment includes
MEXT and JSPS (Japan); ARC and DEST (Australia); NSFC (contract
No.~10175071, China); DST (India); the BK21 program of MOEHRD and the
CHEP SRC program of KOSEF (Korea); KBN (contract No.~2P03B 01324,
Poland); MIST (Russia); MESS (Slovenia); NSC and MOE (Taiwan); and DOE
(USA). The work of Babar are supported by DOE and NSF (USA),
NSERC (Canada), IHEP (China), CEA and CNRS-IN2P3 (France),
BMBF and DFG (Germany), INFN (Italy), FOM (The Netherlands),
NFR (Norway), MIST (Russia), and PPARC (United Kingdom).

\end{acknowledgments}
%

\end{document}